\documentclass[a4paper,11pt]{article}

\usepackage{amssymb}
\usepackage{amsmath}
\usepackage{graphicx}
\usepackage{comment}

\begin{document}

\title{Equation-of-state model for shock compression of hot dense matter}

\author{J.C. Pain\footnote{CEA/DIF, B.P. 12, 91680 Bruy\`eres-Le-Ch\^atel Cedex, France, phone: 00 33 1 69 26 41 85, fax: 00 33 1 69 26 70 94, email: jean-christophe.pain@cea.fr}}

\maketitle


\begin{abstract}
A quantum equation-of-state model is presented and applied to
the calculation of high-pressure shock Hugoniot curves beyond the
asymptotic fourfold density, close to the maximum compression where quantum
effects play a role. An analytical estimate for the maximum attainable compression is proposed. It gives a
good agreement with the equation-of-state model.
\end{abstract}


\section{Introduction}

Maximum pressures reached nowadays by the shock-wave technique in laboratory
laser experiments are as high as hundreds of
megabars and even more for many materials. As the strength of the shock is
varied for a fixed initial state, the
pressure-density final states of the material behind the shock belong to a
curve named shock adiabat or Hugoniot curve. The maximum compression attainable
by a single shock
is finite and occurs for a finite pressure. This phenomenon is due to the
draining of energy in internal degrees of freedom, \textit{via}
ionization and excitation. In this region, the electrons from the ionic cores
are being ionized and the shock
density increases beyond the fourfold density $4\rho_0$ ($\rho_0$ being inital
density), which corresponds to an asymptotic situation where ionization is
completed and the plasma approaches an ideal gas of nuclei and electrons. 
The Hugoniot curve, and therefore the maximum compression, depend on the
equation of state (EOS) of the matter, which, in principle, can be
determined from theory. In practice, the barriers to \textit{ab initio}
calculations are formidable owing to the computational difficulty of solving the
many-body problem. Consequently, it has proven necessary to introduce
simplifying approximations into the governing equations. Many EOS models rely on
the Thomas-Fermi (TF) model
of dense matter \cite{fey}. This approach contains certain essential features in
order
to characterize the material properties but is valid over a limited range of
conditions and provides only a semi-classical description of electrons. However,
at intermediate shock pressures, when the material becomes partially ionized,
the
EOS depends on the precise quantum-mechanical state of the matter, \textit{i.e.}
on the electronic shell structure. 

In Sec. \ref{sec1}, a quantum
self-consistent-field (QSCF) EOS model is presented. The potential and the
bound-electron wavefunctions are determined solving Schr\"odinger equation by a
self-consistent procedure in the density functional theory using a
finite-temperature exchange-correlation potential. Hugoniot curves are
calculated and focus is put on the region where quantum shell effets are
visible, in the vicinity of the maximum compression. In Sec. \ref{sec2}, an
analytical expression for the maximum
compression attainable by a single shock is proposed. It can be applied to any
material from any initial state
except those with gaseous densities. Maximum compression evaluated by this
formula is compared to the one obtained from our QSCF model presented in Sec.
\ref{sec1}.


\section{\label{sec1}Quantum self-consistent-field equation-of-state model}

Hugoniot curves are obtained by resolution of equation

\begin{equation}\label{hug}
(P+P_0)(1/\rho_0-1/\rho)=2(E-E_0),
\end{equation}

which requires the knowledge of pressure $P$ and internal energy $E$ versus
density $\rho$ and temperature $T$ over a wide region of the phase diagram.
Subscript $0$ characterizes the initial state. In the present work, focus is put
on the standard Hugoniot ($P_0$=0, $\rho_0$ equal to solid density and $T_0$=300
K). The adiabatic approximation is
used to separate the thermodynamic functions into electronic and ionic components.
The total pressure of the plasma can be written

\begin{equation}\label{eos1}
P(\rho,T)=P_e(\rho,T)+P_i(\rho,T),
\end{equation}
 
where $P_e(\rho,T)$ and $P_i(\rho,T)$ are respectively the electronic and ionic
contributions to the pressure. The electronic contribution to the EOS stems from
the excitations 
of the electrons due to temperature and compression. Atoms in a plasma can be
idealized by an average atom confined in a Wigner-Seitz (WS) sphere, which
radius $r_{ws}$ is given by solving

\begin{equation}\label{rad}
\frac{4}{3}\pi r_{ws}^3\rho \frac{N_A}{A}a_0^3=1,
\end{equation}

where $a_0$=52.9177208319 10$^{-8}$ cm is the Bohr radius. $N_A$ the Avogadro
number, $\rho$ the matter density in g/cm$^3$, and $A$ the atomic weight in g.
Throughout the paper, all other quantities are in atomic units (a.u.) defined by
$m=\hbar=e=1$, $m$ and $e$ being respectively electron mass and charge. Inside
the sphere, the electron density has the following form \cite{roz2}:

\begin{equation}\label{rel1}
n(r)=\sum_bf_l(\epsilon_b-\mu)\Big|\psi_b(\vec{r})\Big|^2+\frac{\sqrt{2}(k_BT)^{
3/2}}{\pi^2}J_{1/2}\Big(-\bar{V}(r),\bar{\mu}-\bar{V}(
r)\Big)
\end{equation}

where 

\begin{equation}
f_l(x)=\frac{2(2l+1)}{1+\exp[x/k_BT]}
\end{equation}

is the usual Fermi-Dirac population and 

\begin{equation}
J_{n/2}(a,x)=\int_a^{\infty}\frac{y^{n/2}}{1+\exp(y-x)}dy
\end{equation}

is the incomplete Fermi function of order $n/2$. The first term in (\ref{rel1})
is the contribution of bound electrons to the charge
density, while the second term is the free-electron contribution, written in its
semi-classical TF form. The wavefunction $\psi_b$ associated to energy
$\epsilon_b$ of a bound orbital is solution of the Schr\"odinger equation

\begin{equation}\label{sch1}
-\frac{1}{2}\Delta \psi_b+V(r)\psi_b=\epsilon_b\psi_b,
\end{equation}

$V(r)=k_BT\bar{V}(r)$  being the self-consistent potential:

\begin{equation}\label{scf2}
V(r)=-\frac{Z}{r}+\int_0^{r_{ws}}\frac{n(r')}{|\vec{r}-\vec{r'}|}d^3r'+V_{xc}(r)
,
\end{equation}

where $V_{xc}$ is the exchange-correlation contribution, evaluated in the local
density approximation \cite{iye}. Last, the chemical potential $\mu$ is obtained
from the
neutrality of the ion sphere:

\begin{equation}\label{scf3}
\int_0^{r_{ws}}n(r)4\pi r^2dr=Z,
\end{equation}

and $\bar{\mu}=\mu/(k_BT)$. Eqs. (\ref{rel1}), (\ref{sch1}), (\ref{scf2}) 
and (\ref{scf3}) must be solved self-consistently. The electronic pressure $P_e$
\cite{pai,pai2} consists of three
contributions, $P_e=P_{b}+P_{f}+P_{xc}$, where the bound-electron pressure
$P_{b}$ is evaluated at the boundary of the WS sphere using the stress-tensor
formula

\begin{equation}\label{presb}
P_b=\sum_{b}\frac{f_l(\epsilon_b-\mu)}{8\pi
r_{ws}^2}\Big[\Big(\frac{dy_b}{dr}\Big|_{r_{ws}}\Big)
^2+\Big(2\epsilon_b-\frac{1+l+l^2}{r_{ws}^2}\Big)y_b^
2(r_{ws})\Big],
\end{equation}

$y_b$ being the radial part of the wavefunction $\psi_b$ multiplied by $r$. The
free-electron pressure $P_{f}$ reads

\begin{equation}\label{presf}
P_f=\frac{2\sqrt{2}}{3\pi^2}(k_BT)^{5/2}J_{3/2}\Big(-\bar{V}(r_{ws}
),\bar{\mu}-\bar{V}(r_{ws})\Big)\nonumber\\
\end{equation}

and $P_{xc}$ is the exchange-correlation pressure evaluated in the local density
approximation \cite{iye}. The value of bound-electron pressure $P_b$ depends on
the boundary conditions. This comes from the fact that the energy of
an orbital depends on the value of the corresponding wavefunction and on its
derivative at the
boundary of the WS sphere. Table \ref{tab1} represents the energies of the
orbitals of iron (Fe, $Z$=26) obtained if the wavefunction cancels at the
boundary (BC1), if the logarithmic derivative of the wavefunction cancels at the
boundary (BC2) and if the wavefunction behaves like a decreasing exponential at
the boundary (BC3). In this example, the mass density is equal to 34.83 g/cm$^3$
and the
temperature to 115.30 eV. Such conditions belong to the Hugoniot of Fe (see Fig.
\ref{fig2}) calculated from the present QSCF model.
The values of orbital
energies displayed in table \ref{tab1} show that the orbitals calculated with
the three different boundary conditions will not be ionized for the
same value of density. We chose to
perform single-state calculations with boundary condition
(BC3), since it allows the matching of wavefunctions
with Bessel functions outside the WS sphere \cite{pai}, where the potential is
zero. Moreover, condition (BC3) is consistent with the fact that the
bound-electron density vanishes at infinity. The internal energy is

\begin{eqnarray}\label{int}
E_e&=&\sum_kn_k\epsilon_k-\frac{1}{2}\int_0^{r_{ws}}n(r)\int_0^{r_{ws}}
\frac{n(r')}{|\vec{r}-\vec{r'}|}d^3rd^3r'\nonumber\\
&&+E_{xc}-\int_0^{r_{ws}} n(r)V_{xc}(n(r))d^3r,
\end{eqnarray}

where $E_{xc}$ is the exchange-correlation internal energy and $n_k$ the
population of state $k$ (either bound or free). The first term in (\ref{int})
can be expressed by

\begin{equation}
\sum_kn_k\epsilon_k=\sum_bf_l(\epsilon_b,\mu)\epsilon_b+E_{f,k}+E_{f,p}.
\end{equation}

$E_{f,p}$ is the free-electron potential energy

\begin{equation}
E_{f,p}=\frac{\sqrt{2}(k_BT)^{3/2}}{\pi^2}\int_0^{r_{ws}}J_{1/2}\Big(
-\bar{V}(r),\bar{\mu}-\bar{V}(r)\Big)\bar{V}(r)d^3r
\end{equation}

and $E_{f,k}$ the free-electron kinetic energy

\begin{equation}\label{rel2}
E_{f,k}=\frac{\sqrt{2}(k_BT)^{5/2}}{\pi^2}\int_0^{r_{ws}}
J_{3/2}\Big(-\bar{V}(r),\bar{\mu}-\bar{V}(r)\Big)d^3r.
\end{equation}

The ionic contribution to the EOS can be estimated in the ideal-gas
approximation including non-ideality corrections to the thermal motion of
ions. We used an approximation based on the calculation of the EOS of a
one-component
plasma (OCP) by the Monte Carlo method \cite{han}. The ion contribution can be
obtained using a formula based on the Virial theorem:

\begin{equation}\label{ig2}
P_i(\rho,T)=\rho k_BT+\frac{\rho}{3}\Delta E_i(\rho,T),
\end{equation}

where 

\begin{equation}\label{ig3}
\Delta
E_i(\rho,T)=k_BT\;[\Gamma^{3/2}\sum_{i=1}^4\frac{a_i}{(b_i+\Gamma)^{i/2}}-a_1
\Gamma]
\end{equation}

and $E_i(\rho,T)=3k_BT/2+\Delta E_i(\rho,T)$ with $a_1=-0.895929$,
$a_2=0.11340656$, $a_3=-0.90872827$, $a_4=0.11614773$, $b_1=4.666486$,
$b_2=13.675411$, $b_3=1.8905603$ and $b_4=1.0277554$. $\Gamma$ is the plasma
coupling parameter (ratio of ionic Coulomb interaction and thermal kinetic
energy). In the present study, $\Gamma>1$.

The contributions $P_e(\rho,T)$ and $E_e(\rho,T)$ calculated from the QSCF model
are not valid for $T\rightarrow 0$. Therefore, $P_e(\rho,0)$ and $E_e(\rho,0)$
must be substracted from total pressure and internal energy. They are replaced
by $P_c(\rho)$ and $E_c(\rho)$, which constitute the cold curve. The total
pressure reads now

\begin{equation}
P(\rho,T)=P_e(\rho,T)+P_i(\rho,T)-P_e(\rho,0)+P_c(\rho).
\end{equation}

In our model, the cold curve is obtained
either from Augmented Plane Waves (APW) \cite{louc} simulations, or using the
Vinet \cite{vin} universal
EOS. In many situations, Vinet EOS gives realistic results, and its
accuracy is sufficient for most applications involving high pressure and high
temperature situations.

\begin{figure}
\begin{center}
\includegraphics[width=8.6cm,trim=0 0 0 -50]{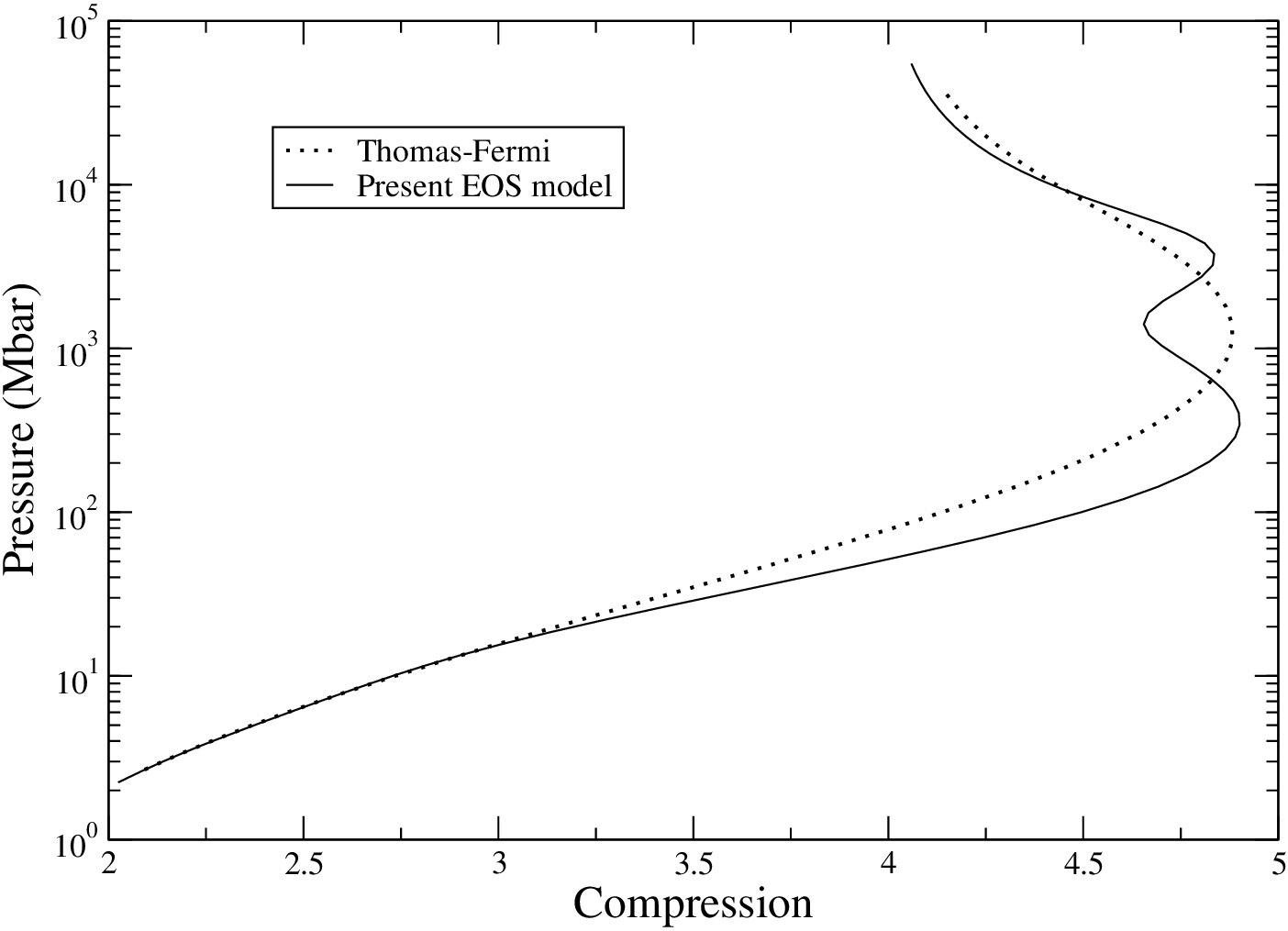}
\end{center}
\caption{Standard Hugoniot curve for Al ($\rho_0=$2.70 g/cm$^3$) calculated from
a model relying on Thomas-Fermi approximation and from our QSCF EOS model.}
\label{fig1}
\end{figure}

\begin{figure}
\begin{center}
\includegraphics[width=8.6cm,trim=0 0 0 -50]{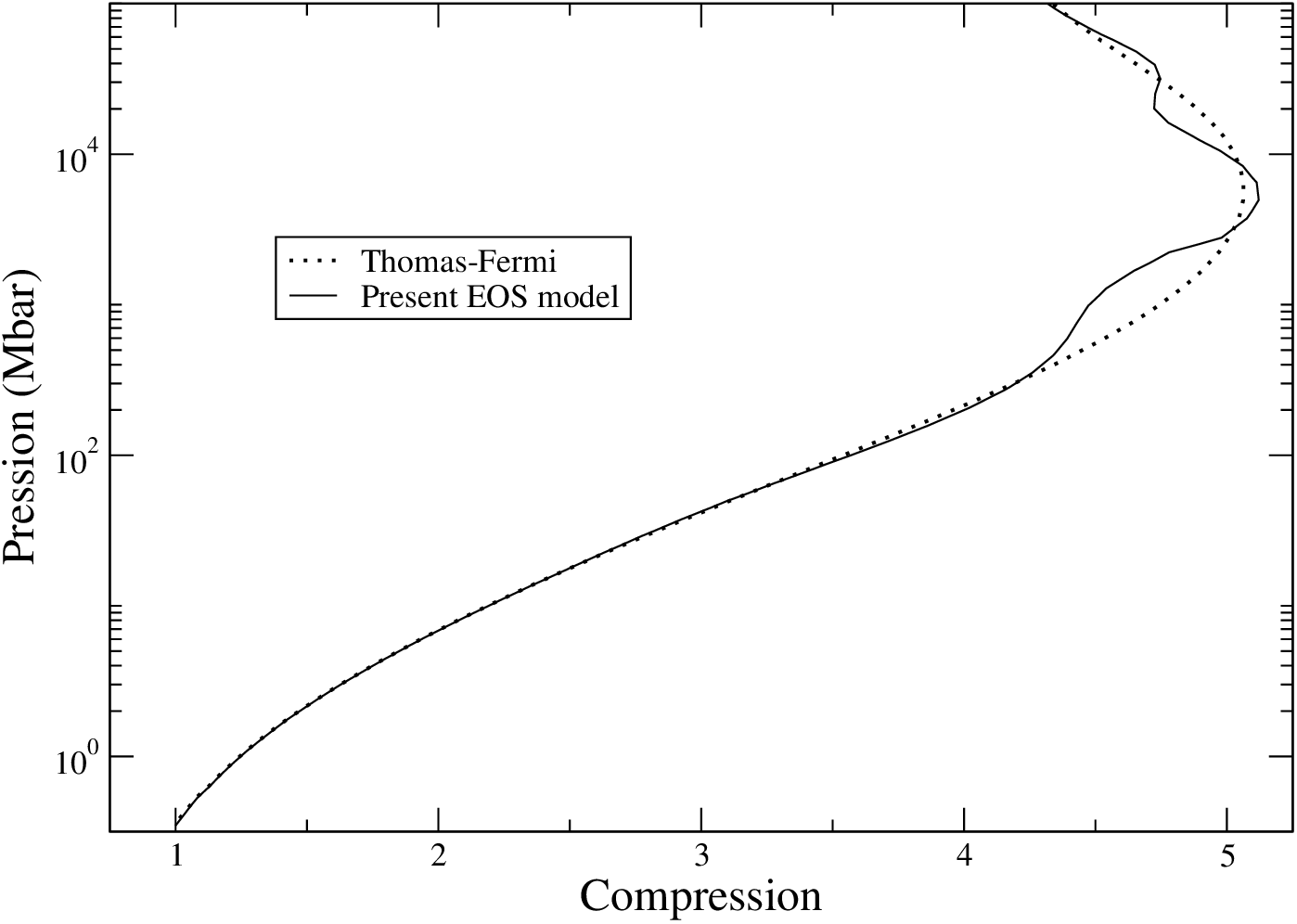}
\end{center}
\caption{Standard Hugoniot curve for Fe ($\rho_0=$7.85 g/cm$^3$) calculated from
a model relying on Thomas-Fermi approximation and from our QSCF EOS model.}
\label{fig2}
\end{figure}

Fig. \ref{fig1} represents the principal Hugoniot curve for Al calculated
from a model relying on Thomas-Fermi approximation and from the present QSCF
model. Fig. \ref{fig2} illustrates the same calculations in the case of Fe.
One can check that the maximum compression is beyond the
ideal-gas asymptote $\rho=4\rho_0$. The oscillations are a
consequence of the competition between the release of energy stocked as internal
energy within the shells and the free-electron pressure. When ionization begins,
the energy of the shock is used mainly to depopulate the relevant shells and the
material is very compressive. However, the pressure of free electrons in
increasing number dominates again and the material becomes more difficult to
compress. The two shoulders in the Hugoniot of Al are due to the successive
ionization of L and K shells respectively. Similarly, the three shoulders in the
Hugoniot of Fe are due to the successive ionization of M, L and K shells
respectively. Table \ref{tab2} illustrates the impact of the choice of the
boundary conditions of the wavefunctions and of the calculation of the ionic
part (perfect ideal gas (IG) with or without OCP corrections
(\ref{ig2}-\ref{ig3})) on the maximum compression rate. As for the
eigen-energies (see table \ref{tab1}), (BC1) gives
the highest value, (BC2) the lowest and (BC3) the intermediate one. The
differences concerning the maximum compression rate appear to be less than 1 \%.
OCP
corrections systematically increase the maximum compression rate.


\section{\label{sec2}Simple evaluation of maximum compression}

The total energy can be written as the sum
of kinetic and potential energies $E_k$ and $E_p$. Neglecting
exchange-correlation contribution for a sake of simplicity, the virial theorem
enables one to relate pressure, kinetic energy and potential energy:

\begin{equation}
3\frac{P}{\rho}=2E_k+E_p.
\end{equation}

Using the Hugoniot relation (\ref{hug}), the compression rate $\eta=\rho/\rho_0$
for the standard Hugoniot ($P_0=0$, $\rho_0$ solid density and $T_0$=300 K) can
be written

\begin{equation}\label{ratemax}
\eta=4+\frac{3}{1+2\frac{E_k-E_{k_0}}{E_p-E_{p_0}}}=4+\frac{3}{1+2\frac{\Delta
E_k}{\Delta E_p}}.
\end{equation}

At high compression, assuming that
$E_k>>E_{k_0}$ and considering that all the electrons have been ionized and that
all electrons have a kinetic energy equal
to the Fermi energy, it is possible to write:

\begin{equation}\label{approx1}
\Delta E_k=Z\frac{1}{2}(3\pi^2Z\rho\frac{N_A}{A})^{2/3}a_0^2.
\end{equation}

At high compression, the excess potential energy can be
estimated as the Coulomb interaction energy of two ionic spheres at close
contact

\begin{equation}\label{approx2}
\Delta E_p=\frac{1}{2}\frac{Z^2}{r_{ws}} \;\;\mbox{and}\;\;
r_{ws}=\Big[\frac{3A}{4\pi N_Aa_0^3}\Big]^{1/3}\rho^{-1/3},
\end{equation}

where $r_{ws}$ is the WS radius defined by Eq. (\ref{rad}). Eqs. (\ref{approx1})
and (\ref{approx2}) are relevant
for a strongly coupled gas of degenerate electrons. Therefore, putting Eqs.
(\ref{approx1}) and (\ref{approx2}) in Eq. (\ref{ratemax}), the maximum
compression rate $\eta_m$ obeys the following equation

\begin{equation}
\eta_m=4+\frac{3}{1+\gamma(\rho_0,Z,A)\eta_m^{1/3}},
\end{equation}

with

\begin{equation}
\gamma(\rho_0,Z,A)=3\pi(2 N_A)^{1/3}a_0\Big(\frac{\rho_0}{ZA}\Big)^{1/3}.
\end{equation}

Intermediate variable $X=\eta_m^{1/3}$ obeys the following fourth order
equation 

\begin{equation}
\gamma X^4+X^3-4\gamma X-7=0.
\end{equation}

The solution is

\begin{equation}\label{eta1}
\eta_m=\Big[\frac{-1+2\epsilon\gamma\sqrt{h(\gamma)}}{{4\gamma}}+\frac{1}{2}
\sqrt{\frac{3}{4\gamma^2}-h(\gamma)+\epsilon\frac{32\gamma^3-1}{4\gamma^3
\sqrt{h(\gamma)}}}\Big]^3
\end{equation}

with $\epsilon=-1$ if $\gamma\le 0.314980262473$ and $\epsilon=+1$ else,

\begin{equation}
h(\gamma)=\frac{1}{4\gamma^2}-\frac{2^{10/3}}{\Delta^{1/3}(\gamma)}+\frac{\Delta
^{1/3}(\gamma)}{2^{1/3}\gamma},
\end{equation}

and 

\begin{equation}
\Delta(\gamma)=-7+16\gamma^3+\sqrt{49+1824\gamma^3+256\gamma^6}.
\end{equation}

\begin{figure}
\begin{center}
\includegraphics[width=8.6cm]{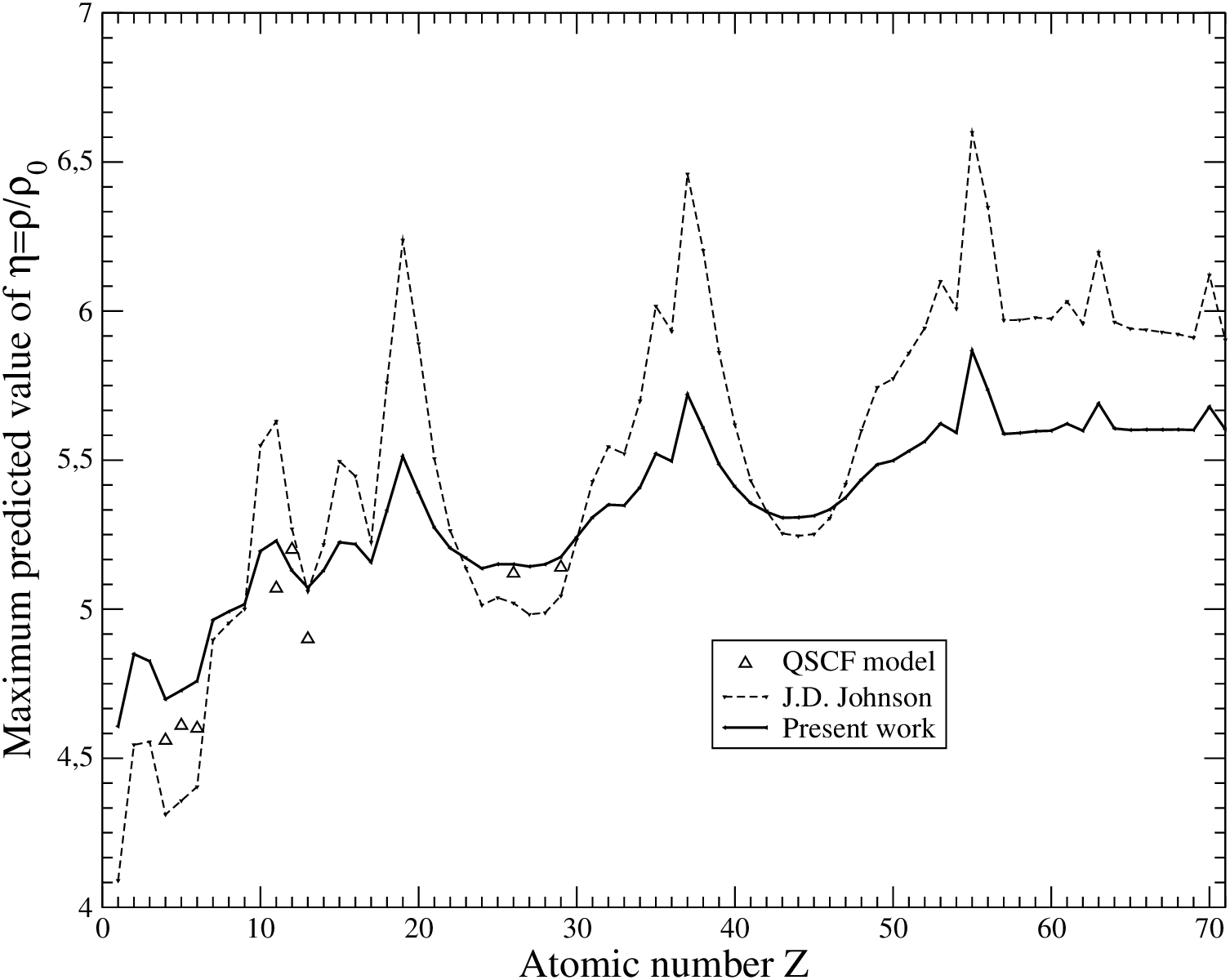}
\end{center}
\caption{Maximum compression rate obtained from \cite{joh}, from our model and
compared to the maximum compression calculated from our QSCF model for Be, B, C,
Na, Mg, Al, Fe and Cu.}
\label{fig3}
\end{figure}

Neglecting cohesive and dissociation energies, and using fits for the total
ionization energies, J.D. Johnson \cite{joh} has proposed an analytical formula
for the maximum compression rate. Fig. \ref{fig3} displays the maximum
compression rate obtained from our analytical formula, from Ref. \cite{joh} and
calculated with our QSCF model for Be, B, C, Na, Mg, Al, Fe and Cu. Exact values
for those particular elements are indicated in table \ref{tab3}. It first
confirms the fact that the maximum compression is always smaller than 7
\cite{joh}, and strongly dependent on the density $\rho_0$. The most important
point is that the maximum compression obtained with formula (\ref{eta1}) is very
close to the one obtained from our QSCF model. This means that Eq. (\ref{eta1}),
which does not rely on quantum mechanics, gives a good agreement with a quantum
EOS model. At first sight, it seems difficult to decide in an unequivocal way
which
approach, between formula (\ref{eta1}) and Ref. \cite{joh}, is the most
accurate. However, our estimate (\ref{eta1}) does not rely
on a particular form of the EOS and is fully analytical, since it does not
require
any data concerning the ionization energies, as in \cite{joh}. On the contrary
to the formula proposed in \cite{joh}, the maximum compression predicted by our
analytical model is higher for Fe than for Al, which is consistent with the
results
presented in \cite{roz1}. However, in \cite{roz1} the authors notice that the
maximum compression seems to increase with $Z$, even if they confess that they
have no explanation for that. We believe this is only a global tendency, and
Fig. \ref{fig3} shows that the maximum compression, evaluated
either from Ref. \cite{joh} or from Eq. (\ref{eta1}), does not vary
monotonically with atomic number $Z$.


\section{Conclusion}

A quantum equation-of-state model was presented. It consists in a
self-consistent calculation of the electronic structure. Bound electrons are
treated in the framework of quantum mechanics and bound-electron pressure is
evaluated using the stress-tensor formula. Free electrons are described in the
Thomas-Fermi approximation. Exchange-correlation effects at finite temperature
are taken into account. The ionic part is described in the ideal-gas
approximation with non-ideality corrections from the one-component plasma. It
was shown that such a model is well suited for the computation of Hugoniot
curves and exhibits oscillations due to the ionization of successive shells. A
simple analytical estimate was proposed for the maximum compression attainable
by a single shock, which is crucial for the diagnostic of high-pressure
laser-induced shock waves. It relies on an expression for the increase of
kinetic energy assuming degenerate electrons and the increase of potential
energy is evaluated as Coulomb interation energy between hard spheres.
Furthermore, the maximum compression obtained from this formula is in good
agreement with the one calculated from our quantum EOS model.


\clearpage

\begin{table}
\begin{center}
\begin{tabular}{|c|c|c|c|} \hline \hline
 Orbital & $\epsilon^{(BC1)}$ (a.u.) & $\epsilon^{(BC2)}$ (a.u.) &
$\epsilon^{(BC3)}$ (a.u.) \\
\hline \hline
 1s & $-256.72931$ & $-257.32658$ & $-256.98981$\\
 2s & $-29.891050$ & $-30.363675$ & $-30.094431$\\
 2p & $-25.511960$ & $-26.002778$ & $-25.723564$\\
 3s & $-2.3491653$ & $-2.8962546$ & $-2.5844217$\\
 3p & $-1.0321504$ & $-1.6990494$ & $-1.3273471$\\\hline \hline
\end{tabular}
\end{center}
\caption{Energies of orbitals of Fe at $\rho$=34.83 g/cm$^3$ and T=115.30 eV for
three boundary conditions: (BC1): the wavefunction is zero at the boundary,
(BC2): the logarithmic derivative of the wavefunction is zero at the boundary
and (BC3): the wavefunction behaves like a decreasing exponential at the
boundary.}\label{tab1}
\end{table}

\begin{table}
\begin{center}
\begin{tabular}{|c|c|c|} \hline \hline
 Boundary cond. & Ionic part & Max. compression rate \\\hline \hline
 (BC1) & IG & $4.908$ \\
 (BC2) & IG & $4.891$ \\
 (BC3) & IG & $4.901$ \\
 (BC3) & IG+OCP corr. & $4.931$ \\ \hline \hline
\end{tabular}
\end{center}
\caption{Impact of the boundary conditions of the bound-electron wavefunctions
and of the treatment of ions on the maximum compression for Al.}\label{tab2}
\end{table}

\begin{table}
\begin{center}
\begin{tabular}{|c|c|c|c|c|c|} \hline \hline
 Element & $Z$ & $\eta_m[QSCF]$ & $\eta_m[$Formula $(\ref{eta1})]$ &
$\eta_m[$Ref. $\cite{joh}]$ \\
\hline \hline
 Beryllium & $ 4$ & $4.56$ & $4.70$ & $4.31$\\
 Boron & $ 5$ & $4.61$ & $4.73$ & $4.36$\\
 Carbon & $ 6$ & $4.60$ & $4.76$ & $4.40$\\
 Sodium & $11$ & $5.07$ & $5.23$ & $5.63$\\
 Magnesium & $12$ & $5.20$ & $5.13$ & $5.27$\\ 
 Aluminum & $13$ & $4.90$ & $5.07$ & $5.06$\\
 Iron & $26$ & $5.12$ & $5.15$ & $5.02$\\
 Copper & $29$ & $5.14$ & $5.18$ & $5.04$\\\hline \hline
\end{tabular}
\end{center}
\caption{Maximum compression rate for different elements calculated from our
QSCF equation-of-state model, from our analytical formula (\ref{eta1}) and from
an estimate published in Ref. \cite{joh}.
}\label{tab3}
\end{table}

\end{document}